\documentclass[twocolumn,aps,showpacs,eqsecnum]{revtex4}

\usepackage{latexsym}
\usepackage{amsmath}
\usepackage{amssymb}
\usepackage{epsfig}

\newcommand{\bra}[1]{\langle #1|}
\newcommand{\ket}[1]{| #1 \rangle }
\newcommand{\Bra}{\langle}

\newcommand{\C}{{\rm \Pi}}

\newcommand{\Hs}{{\cal H}}
\newcommand{\one}{{I}}
\newcommand{\sym}{{P_{sym}}}
\newcommand{\complex}{{\mathbb C}}

\begin{document}
\title{Unambiguous comparison of ensembles of quantum states}
\author{Michal Sedl\'ak$^{1,3}$,
M\'ario Ziman$^{1,2}$,
Vladim\'\i r Bu\v zek$^{1,3}$,
and Mark Hillery$^4$
}
\address{
$^{1}$Research Center for Quantum Information, Slovak Academy of Sciences, D\'ubravsk\'a cesta 9, 845 11 Bratislava, Slovakia
}
\address{
$^{2}$Faculty of Informatics, Masaryk University, Botanick\'a 68a, 602 00 Brno, Czech Republic
}
\address{
$^{3}${\em Quniverse}, L{\'\i}\v{s}\v{c}ie \'{u}dolie 116, 841 04 Bratislava, Slovakia
}
\address{
$^4$Department of Physics and Astronomy, Hunter College
of the City University of New York, 695 Park Avenue, New York, NY 10021, USA
}

\begin{abstract}
We present a solution of the problem of the optimal unambiguous comparison of two ensembles of unknown quantum
states $\ket{\psi_1}^{\otimes k}$ and $\ket{\psi_2}^{\otimes l}$. We consider two cases: 1) The two unknown states
$\ket{\psi_1}$ and $\ket{\psi_2}$ are arbitrary states of qudits. 2) Alternatively, they are coherent states of a harmonic
oscillator. For the case of coherent states we propose a simple experimental realization of the optimal
``comparison'' machine composed of a finite number of beam-splitters and a single photodetector.
\end{abstract}

\maketitle

\section{Introduction}

In the classical world it is relatively easy to compare (quantitatively,
or qualitatively) features of physical systems and to conclude with
certainty whether the systems possess the same
properties, or not.
On the other hand, the statistical nature of the quantum theory
restricts our ability to provide deterministic conclusions/predictions
even in the simplest experimental situations. Therefore comparison
of quantum states is totally different compared to classical situation.
To be specific, let us consider that we are given two quantum systems
of the same physical origin (e.g., two photons) and our task is to conclude
whether these two photons have been prepared in the same polarization state.
That is, we want to compare the two states and we want to know whether
they are identical or not. Given the fact that we have just a single
copy of each state the scenario according to which we first measure each state does not work. For that we would need an infinite
ensemble of identically prepared systems. The solution to the problem of comparison of quantum states has been proposed by
Barnett {\it et al.} \cite{jex}: Within quantum realm we can compare
two states, but there is a price to pay. For instance, one cannot conclude
with certainty that two systems are in the same {\em pure }state or not, except
for the case when the set of possible
 pure states is linearly independent \cite{chefles}. The {\em unambiguous} state comparison as introduced by Barnett {\it et al.} is a
positive-operator-value-measure (POVM) measurement that has two possible outcomes associated with the two answers: the two states
are different, or outcome of the measurement corresponds to an inconclusive answer.
Moreover, the existence of the negative answer strongly depends on the particular
quantum states $\varrho_1,\varrho_2$ in the following sense. To give the unambiguous conclusion that the states are different it is
necessary to restrict ourselves to states, which have distinct supports \cite{kleinman}.
In the quantum comparison problem as discussed by Barnett, Chefles, Jex and Andersson  \cite{jex,andersson,chefles}
it is assumed that the unknown states are pure  and only a single copy of each of them is available.

The aim of the present paper is to find the optimal unambiguous state comparison
procedure in the case we have more copies of the two quantum states which we want to compare.
Throughout the paper we assume that the compared states are pure and that they belong
to a $d$-dimensional Hilbert space $\Hs$. The dimensionality of the Hilbert space in known, otherwise
no further information about the states is available. In the case of (semi)-infinite dimensional Hilbert space $\Hs_{\infty}$
(corresponding to a harmonic oscillator) we restrict our investigation to a specific case, when we a priori know that the
two states to be compared are coherent states. What is not known are their complex amplitudes.
Our goal will be to design an optimal quantum comparison machine.

As in the case of only one copy per each of the two compared states it is not possible to unambiguously conclude
that the compared states are the same. Thus, the positive operator-valued
measure (POVM) describing the measurement apparatus \cite{nielsen} will have
only two measurement elements $\C_0$ indicating the failure of the
measurement and $\C_1=I-\C_0$ unambiguously showing that the compared states are different.

In the paper we will derive the optimal multi-copy comparator for
general pure states (Sec.~II) and for coherent states (Sec.~III).
In both cases we will investigate the behavior of the success probability
as a function of  the number of copies $k$ and $l$ of the two states. Moreover, we will
propose a relatively simple experimental setup realizing the
comparison of coherent states.

\section{Comparison of states of finite-dimensional systems}
Let us consider that we have $k$ copies of the first unknown state
(further denoted as $\ket{\psi_1}$) and $l$ copies of the second
unknown state (denoted as $\ket{\psi_2}$). Our task is to
either unambiguously conclude that the states $\ket{\psi_1},\ket{\psi_2}$
are different, or to admit that we cannot give a definite answer whether they are identical or different.
The optimal measurement that would allow us to implement  this task  follows from
the work by Chefles {\it et al.} \cite{chefles} who analyzed the problem from a more general
perspective. They have discussed theoretical framework which allow one to evaluate
the probability of success. In our work  we provide a short derivation of the optimal measurement and explicitly
evaluate the probability of success in such measurement. The aforementioned derivation will guide us in our quest
for finding the optimal measurement that would allow us to compare coherent states.

In order to construct the desired POVM for the state comparison we first introduce
the (no-error) condition that guarantees that whenever
we obtain the result $\C_1$ we can conclude that the states were
indeed different:
\begin{eqnarray}
\forall \ket{\psi}\in \Hs, \quad
Tr[\C_1 (\ket{\psi} \bra{\psi})^{\otimes k+l}]=0\, .
\label{noerror1}
\end{eqnarray}
Integrating uniformly over all pure states $S_d=\{\ket{\psi}\in\Hs\}$
we obtain an equivalent no-error condition that reads
\begin{eqnarray}
0&=&\int_{S_d}d\psi
{\rm Tr}\Big[\C_1(\ket{\psi} \bra{\psi})^{\otimes k+l}\Big]
={\rm Tr}[\C_1 \Delta] \, ,
\label{nerorint}
\end{eqnarray}
where
\begin{eqnarray}
\Delta=\int_{S_d}d\psi (\ket{\psi} \bra{\psi})^{\otimes k+l}
=\frac{1}{\binom{k+l+d-1}{d-1}} \sym ,
\label{delta}
\end{eqnarray}
and $\sym$ is the projector onto a completely symmetric subspace of
$\Hs^{\otimes (k+l)}$ and $d$ is the dimension of the Hilbert space.
The derivation of the formula (\ref{delta}) can be found for example
in the paper of Hayashi {\it et al.} \cite{hayashi1}.

Because of the positivity of the operators $\C_1$ and $\Delta$ the equation
(\ref{nerorint}) implies that these two operators have orthogonal supports.
Hence the largest possible support the operator
$\C_1$ can have is the orthogonal complement to the support of $\Delta$.
The support of the projector $\one-\sym$ is therefore the largest possible
support of $\C_1$. The optimal measurement must maximize
the average success probability $\overline{P(k,l)}$
of revealing the difference between the states that are launched into the comparator
\begin{eqnarray}
\overline{P(k,l)}&=&\int_{S_d}\int_{S_d}d\psi_1 d\psi_2
P(\ket{\psi_1},\ket{\psi_2}), \label{mpr} \\
\nonumber
P(\ket{\psi_1},\ket{\psi_2})&=&{\rm Tr}
[\C_1 (\ket{\psi_1}\bra{\psi_1})^{\otimes k}
\otimes(\ket{\psi_2}\bra{\psi_2})^{\otimes l}],
\end{eqnarray}
while keeping the positivity ($0\leq \C_1 \leq\one$)
and the no-error conditions satisfied. Combining these two conditions
on the support of $\C_1$ (for details see Ref.~\cite{chefles}) we obtain $\C_1=\one-\sym$.
Thus the optimal state comparison of $k$ and $l$ copies of a pair of an unknown pure states is
accomplished by the following projective measurement
\begin{eqnarray}
\nonumber
\C^{opt}_0&=&\sym ,\\
\C^{opt}_1&=&\one-\sym \, .
\label{cmopt}
\end{eqnarray}

In what follows we calculate the probability of revealing the difference of the
states $\ket{\psi_1}$, $\ket{\psi_2}$ measured by the optimal comparator,
i.e.
\begin{eqnarray}
\nonumber
P(\ket{\psi_1},\ket{\psi_2})&=&{\rm Tr}[(\one-\sym)\ket{\Psi}\bra{\Psi}] \\
&=&1-\Bra{\Psi}\ket{\Psi_S}\, ,
\label{ovlp1}
\end{eqnarray}
where
$\ket{\Psi}\equiv\ket{\psi_1}^{\otimes k}\otimes\ket{\psi_2}^{\otimes l}$
and
\begin{eqnarray}
\ket{\Psi_S}\equiv \sym\ket{\Psi}=
\frac{1}{(k+l)!}\sum_{\sigma\in S(k+l)} \sigma(\ket{\Psi}) \, .
\label{psis}
\end{eqnarray}
In the above formulas we denoted by $S(n)$ a group of permutations
of $n$ elements and $\sigma(\ket{\Psi})$ denotes the state $\ket{\Psi}$
in which subsystems
have been permuted via the permutation $\sigma$. For example,
a permutation $\nu_k$ exchanging only $k$-th and $(k+1)$-th position
acts as
\begin{eqnarray}
\nu_k(\ket{\Psi})=\ket{\psi_1}^{\otimes k-1}
\ket{\psi_2}\ket{\psi_1}\ket{\psi_2}^{\otimes l-1}\, .
\end{eqnarray}
The state $\ket{\Psi}$ has $n$ subsystems defining $n$ positions,
which are interchanged by the permutation $\sigma$.
Let us denote by $N_1$ the subset of the first $k$ positions
(originally copies of $\ket{\psi_1}$) and by $N_2$ the remaining
$l$ positions (originally occupied by systems in the state $\ket{\psi_2}$).
For our purposes it will be useful to characterize each permutation
$\sigma\in S(k+l)$ by the number of positions $m$
in the subset $N_1$ occupied by subsystems originated from
the subset $N_2$. Literally, $m(\sigma)$ is the number of states $\ket{\psi_2}$
moved into the first $k$ subsystems ($N_1$)
by the permutation $\sigma$ acting on the state $\ket{\Psi}$.
Using this number we can write
\begin{eqnarray}
\Bra{\Psi}\ket{\sigma(\Psi)}=|\Bra{\psi_1}\ket{\psi_2}|^{2m(\sigma)}\, .
\end{eqnarray}
For instance,
\begin{eqnarray}
\bra{\Psi}\nu_k(\ket{\Psi})&=&\bra{\psi_1}^{\otimes k}\bra{\psi_2}^{\otimes l}\ket{\psi_1}^{\otimes k-1}\ket{\psi_2}\ket{\psi_1}\ket{\psi_2}^{\otimes l-1}\nonumber\\
&=&|\Bra{\psi_1}\ket{\psi_2}|^{2m(\nu_k)}=|\Bra{\psi_1}\ket{\psi_2}|^{2}
\, .\nonumber
\end{eqnarray}

In order to evaluate the scalar product
\begin{eqnarray}
\Bra{\Psi}\ket{\Psi_S}
&=&\frac{1}{(k+l)!}\sum_{\sigma \in S(k+l)} \bra{\Psi}\sigma(\ket{\Psi}). \label{sump1}
\end{eqnarray}
we need to calculate the number of permutations $C_m$ with the same value
$m=m(\sigma)$. For each permutation $\sigma$ there are exactly $k!l!$
permutations
leading to the same state $\sigma(\ket{\Psi})$. The number of different
quantum states $\sigma_1(\ket{\Psi}), \sigma_2(\ket{\Psi}), \ldots$ having
the same overlap $|\Bra{\psi_1}\ket{\psi_2}|^{2m}$ with the state
$\ket{\Psi}$ (i.e. the same $m$) is $\binom{k}{m}\binom{l}{m}$. This is because each such state
is fully specified by enumerating $m$ from the first $k$ subsystems to which
$\ket{\psi_2}$ states were permuted and by enumerating $m$ from the last $l$
subsystems to which $\ket{\psi_1}$ states were moved.
To sum up our derivation, we have $C_m=k!l!\binom{k}{m}\binom{l}{m}$, and
consequently Eq.~(\ref{sump1}) can be rewritten as
\begin{eqnarray}
\label{sump2}
\Bra{\Psi}\ket{\Psi_S}&=&
\sum_{m=0}^{\min(k,l)}\frac{\binom{k}{m}\binom{l}{m}}{\binom{k+l}{k}}
|\Bra{\psi_1}\ket{\psi_2}|^{2m}\, .
\end{eqnarray}
The optimal probability reads
\begin{eqnarray}
\label{prpure}
P(\ket{\psi_1},\ket{\psi_2})=1-
\sum_{m=0}^{\min(k,l)}\frac{\binom{k}{m}\binom{l}{m}}{\binom{k+l}{k}}
|\Bra{\psi_1}\ket{\psi_2}|^{2m}\, .
\end{eqnarray}
The average probability is calculated in Appendix A and results in
the following formula
\begin{eqnarray}
\overline{P(k,l)}=1-\frac{\dim(\Hs^{\otimes k+l}_{sym})}{\dim(\Hs^{\otimes k}_{sym})\dim(\Hs^{\otimes l}_{sym})}, \label{dimcnt}
\end{eqnarray}
where $\Hs^{\otimes k}_{sym}$ stands for a completely symmetric subspace
of $\Hs^{\otimes k}$. Thus, we see that the success rate is essentially
given by one minus the ratio of dimensionality of the failure subspace
to the dimension of the potentially occupied space.

\subsection{Additional copy of an unknown state}
Next we will analyze properties
of $P(\ket{\psi_1},\ket{\psi_2},k,l)$. In particular, we will study how it behaves as a function
of the number $k,l$ of available copies
of the two compared states. We are going to confirm
a very natural expectation that any additional copy of one of the
compared states always increases the probability of success.
Stated mathematically, it suffices to prove that
\begin{eqnarray}
P(\ket{\psi_1},\ket{\psi_2},k+1,l) \geq P(\ket{\psi_1},\ket{\psi_2},k,l),
\label{rast1}
\end{eqnarray}
since $P(\ket{\psi_1},\ket{\psi_2},k.l)$ is symmetric with respect to $k,l$.
For $k\ge l$
\begin{eqnarray}
\delta &\equiv& P(\ket{\psi_1},\ket{\psi_2},k+1,l) - P(\ket{\psi_1},\ket{\psi_2},k,l)\nonumber\\
&=&\frac{1}{\binom{k+l}{k}}\sum_{m=0}^{\min(k,l)}\left(1-\frac{(k+1)^2}{(k+1-m)(k+l+1)}\right)\nonumber\\
& \ &\times\binom{k}{m}\binom{l}{m} |\Bra{\psi_1}\ket{\psi_2}|^{2m}\, .
\end{eqnarray}
For $k<l$ the additional term
$-|\Bra{\psi_1}\ket{\psi_2}|^{2k+2}\binom{k+l+1}{k+1}/\binom{l}{k+1}$
appears in the expression for $\delta$, however it is possible to proceed
in the same way in both cases. We can think of $\delta$ as being
a polynomial in $x\equiv |\Bra{\psi_1}\ket{\psi_2}|^2$, which vanishes
for $x=1$, because $P(\ket{\psi},\ket{\psi})=0$. The
coefficients $a_m$ of the polynomial $\delta=\sum_m a_m x^m$ are
nonnegative for $m\leq(k+1)l/(k+l+1)$ and negative otherwise.
Therefore, we can apply the Lemma from Appendix B to conclude that
$\delta(x)\geq 0$ for $x\in[0,1]$, which is equivalent to Eq.(\ref{rast1}).
We have proved that for any pair of compared states the additional copies
of the states improve the probability of success, so the statement holds
also for the average success probabilities, i.e.
\begin{eqnarray}
\overline{P(k+1,l)} \geq \overline{P(k,l)}\, .
\end{eqnarray}

\subsection{Optimal choice of resources}
Now we consider the situation when the total number $N$ of copies of the two states
is fixed, i.e. $N=k+l$. Our aim is to maximize the success probability with
respect to the splitting of the $N$ systems into $k$ copies of the state
$\ket{\psi_1}$ and $l$ copies of the state $\ket{\psi_2}$. In order to find the solution to this problem we
prove the following inequality
\begin{eqnarray}
\nonumber
\Lambda&\equiv& P(\psi_1,\psi_2,k+1,N-k-1)
-P(\psi_1,\psi_2,k,N-k)\\
&\geq& 0 \quad {\rm for}\quad k\leq \lfloor N/2 \rfloor\, ,
\label{rast2}
\end{eqnarray}
where $\lfloor a \rfloor$ indicates the floor function, i.e. the integer part
of the number. The previous inequality automatically implies $\Lambda\leq 0$
for $k>\lfloor N/2 \rfloor$, because $P(\psi_1,\psi_2,k,l)$
is symmetric in $k$ and $l$. Therefore, this would mean that the optimal
value is $k=\lfloor N/2 \rfloor$.

Thus, to complete the proof it is sufficient to confirm the validity
of Eq. (\ref{rast2}). This is done in the same way as for Eq.~(\ref{rast1})
i.e. by looking on $\Lambda$ as on a polynomial in
$x\equiv |\Bra{\psi_1}\ket{\psi_2}|^2$ and showing that the assumptions
of the Lemma from Appendix B hold.

Hence, given the total number $N$ of copies it is most optimal to have half of them in the state
 $\ket{\psi_1}$ and the other half in the state $\ket{\psi_2}$. In this case
the average probability of success
\begin{eqnarray}
\max_k\overline{P(k,N-k)}=\lfloor N/2 \rfloor \,
\end{eqnarray}
is maximized.

\begin{figure}
\begin{center}
\includegraphics[width=8cm]{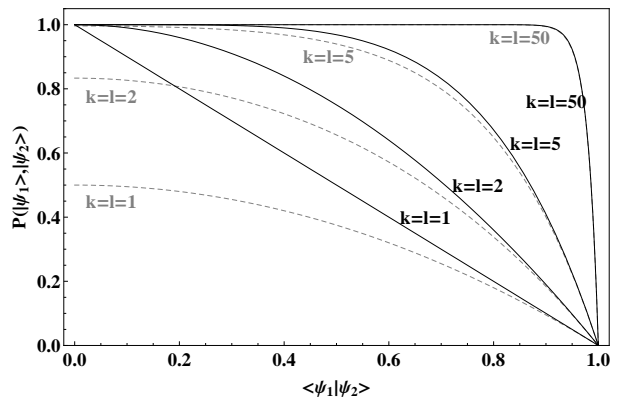}
\caption{The probability of revealing the difference between the compared states $\ket{\psi_1}$, $\ket{\psi_2}$.
The gray dashed lines are valid for the optimal state comparison among all pure states. Each line corresponds to a different number of copies of
the compared states. The solid black lines indicate the performance of the optimal comparison if we are restricted to coherent states only.}
\label{ppvo}
\end{center}
\end{figure}

\begin{figure}
\begin{center}
\includegraphics[width=8cm]{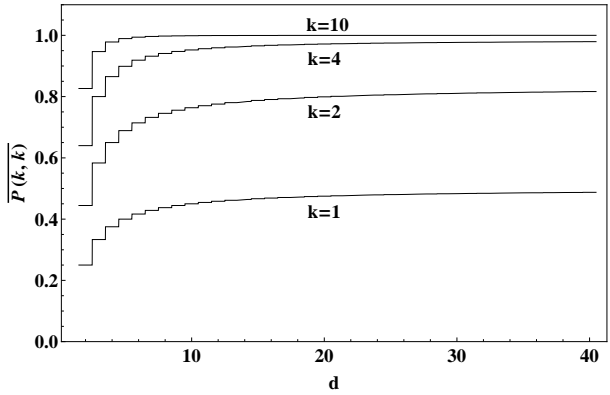}
\caption{The mean probability of  the detection of a difference between the compared states $\ket{\psi_1}$, $\ket{\psi_2}$ as a function of the dimension of
the Hilbert space of the compared systems.}
\label{mpd}
\end{center}
\end{figure}

More quantitative insight into the behavior of $P(\ket{\psi_1},\ket{\psi_2})$
and $\overline{P(k,k)}$ is presented in Figs. (\ref{ppvo}) and (\ref{mpd}).
The figure (\ref{ppvo}) illustrates that the more copies of the compared
states we have and the smaller is their overlap, the higher is the probability
of revealing the difference between the states. The overlap of a pair of randomly
chosen states decreases with the dimension of $\Hs$. Therefore
the mean probability $\overline{P(k,k)}$ for a fixed number of copies $k$ grows
with the dimension $d$. This fact is documented in Fig. (\ref{mpd}).

\subsection{Comparison with large number of copies}

Let us now study the situation when $k=1$ and $l\rightarrow\infty$.
In this case the sum in Eq.~(\ref{prpure})
has only two terms, which can be easily evaluated to obtain
\begin{eqnarray}
P(\ket{\psi_1},\ket{\psi_2})&=&\lim_{l\rightarrow\infty}
\left( 1-\frac{1+l|\Bra{\psi_1}\ket{\psi_2}|^{2}}{l+1}\right)= \nonumber\\
&=&1-|\Bra{\psi_1}\ket{\psi_2}|^{2} \, .
\end{eqnarray}
In this limit the same probability of success can be reached also by
a different comparison strategy. We can first use the state reconstruction
techniques to precisely determine the state $\ket{\psi_2}$ and then
by projecting the remaining $\ket{\psi_1}$ state onto
$\one-\ket{\psi_2}\bra{\psi_2}$ reveal the difference between the states.

For the limit, where the number of both compared states goes
to infinity simultaneously ($k=l\rightarrow\infty$), from
 Eq. (\ref{dimcnt}) we recover
for any finite $d$ the classical behavior i.e.
\begin{eqnarray}
\lim_{k\rightarrow\infty}\overline{P(k,k)}=1 \, .
\end{eqnarray}

Therefore we can conclude that larger the number of the copies $k$ and $l$ of the two states
higher the probability to determine that the two states are different is. In the limit
$k=l\rightarrow\infty$ we essentially with a classical comparison problem.

\section{Comparison of coherent states}

In any quantum information processing the prior knowledge about
the system in which information is encoded plays an important role.
The most explicit example one can name is the state estimation when the prior
knowledge about the state is crucial.
In what follows we will analyze the quantum state comparison
and instead of assuming that the two compared states are totally arbitrary we will
restrict a class of possible states. To be more specific, we will consider a harmonic oscillator
and we focus our attention on  comparison of coherent states.

Coherent states \cite{coherent} are defined as eigenstates of the
annihilation operator
$a$ (acting on ${\cal H}_\infty$) associated with eigenvalues taking arbitrary
value in the complex plane, i.e. the set of coherent states is defined as
\begin{eqnarray}
S_{\rm coh}=\{ \ket{\alpha}\in\Hs_{\infty}:\quad \alpha\in\complex \, , \quad
a\ket{\alpha}=\alpha\ket{\alpha}\}\, .
\end{eqnarray}

Our next task is two-fold: Firstly we introduce an optimal protocol for
comparison of two coherent states.
Secondly we  propose an experimental realization of the optimal
coherent states comparator. Following the same line of reasoning as
in the previous section the measurement operator $\C^{\rm coh}_1$
unambiguously revealing that the coherent states
($k$ copies of state $\ket{\alpha_1}$
and $l$ copies of the state $\ket{\alpha_2}$) are different must
obey the following ``no-error'' conditions
\begin{eqnarray}
{\rm Tr}[\C^{\rm coh}_1 (\ket{\alpha} \bra{\alpha})^{\otimes k+l}]=0
\quad\forall\, \ket{\alpha}\in S_{\rm coh}\, ,
\label{noerror2}
\end{eqnarray}
or equivalently
\begin{eqnarray}
0&=&\int_{S_{\rm coh}} d\alpha
{\rm Tr}\Big[\C^{\rm coh}_1\ket{\alpha}\bra{\alpha}^{\otimes k+l}\Big]
={\rm Tr}[\C^{\rm coh}_1 \Delta]\, ,
\label{nerorint1}
\end{eqnarray}
where $d\alpha$ is an arbitrary positive measure such that its support
contains all coherent states.

Since the operators $\Pi^{\rm coh}_1$ and $\Delta$ are positive, the
identity ${\rm Tr}[\Pi_{\rm coh}\Delta]=0$ implies that their supports
are orthogonal. As before (in the case of all pure states) it is optimal
to choose $\Pi_1^{\rm coh}$ to be  a projector onto the orthocomplement
of the support of $\Delta$. Denoting by $\Delta_{\rm coh}^N$
the projector onto the support of $\Delta$ we can write
$\Pi_1^{\rm coh}=I-\Delta_{\rm coh}^N$. As it is shown in Appendix C
using a properly normalized Lebesgue measure on a complex plane
we can write
\begin{equation}
\Delta=\frac{N}{\pi}\int_{\complex} d\alpha \ket{\alpha}\bra{\alpha}^{\otimes N}
=\Delta_{\rm coh}^{N}\, .
\label{coh_proj}
\end{equation}

Consider $\ket{\Psi}=\ket{\alpha_1}^{\otimes k}\otimes
\ket{\alpha_2}^{\otimes l}$
to be a general input state of the coherent-state comparison
machine. Using the Eq.(\ref{coh_proj})  
we obtain the following expression for the
success probability $P(\ket{\alpha_1},\ket{\alpha_2})$
\begin{eqnarray}
P(\ket{\alpha_1},\ket{\alpha_2})&=&
{\rm Tr}\Big[\C^{\rm coh}_1\ \ket{\Psi}\bra{\Psi}\Big]=
1-\bra{\Psi}\Delta_{\rm coh}^{k+l}\ket{\Psi}\nonumber\\
&=&1-\frac{k+l}{\pi}\int_\complex d\beta |\Bra{\alpha_1}\ket{\beta}|^{2k} |\Bra{\alpha_2}\ket{\beta}|^{2l} \nonumber\\
&=&1-\frac{k+l}{\pi}\int_\complex d\beta e^{-k|\alpha_1-\beta|^2 - l|\alpha_2-\beta|^2 }\nonumber\\
&=&1-\frac{k+l}{\pi}\ e^{-\frac{kl}{k+l}|\alpha_1-\alpha_2|^2}\nonumber\\
&\ &\times \int_\complex d\beta e^{-\Big|\sqrt{k+l}\beta-\frac{1}{\sqrt{k+l}}(k\alpha_1+l\alpha_2)\Big|^2 }\nonumber\\
&=&1-e^{-\frac{kl}{k+l}|\alpha_1-\alpha_2|^2},
\end{eqnarray}
where we used the following modification of the rectangular identity
\begin{eqnarray}
&k&|\alpha_1-\beta|^2+l |\beta-\alpha_2|^2  \nonumber\\
&=&\Big|\sqrt{k+l}\beta- \frac{k\alpha_1+l \alpha_2}{\sqrt{k+l}}\Big|^2+\frac{kl}{k+l}|\alpha_1-\alpha_2|^2.
\nonumber
\end{eqnarray}

\subsection{Optical setup for unambiguous comparison of
coherent states}
In this subsection we will describe an optical realization
of an unambiguous coherent-states comparator that achieves the optimal
value of the success probability (see above). The experimental setup
we are going to propose will
consist of several beam-splitters and only a single photodetector.
A beam-splitter acts on a pair of coherent states in a very convenient
way, in particular, the output beams remain unentangled and coherent, i.e.
\begin{eqnarray}
\ket{\alpha}\otimes\ket{\beta}\mapsto
\ket{\sqrt{T}\alpha+\sqrt{R}\beta}\otimes
\ket{-\sqrt{R}\alpha+\sqrt{T}\beta}\, ,
\label{beamtransf}
\end{eqnarray}
where $T,R$ stand for transmissivity and reflectivity,
respectively, and $T+R=1$. The aforementioned property of the beam-splitter
transformation enables us to consider each of its outputs separately.

\begin{figure}
\begin{center}
\includegraphics[width=9cm]{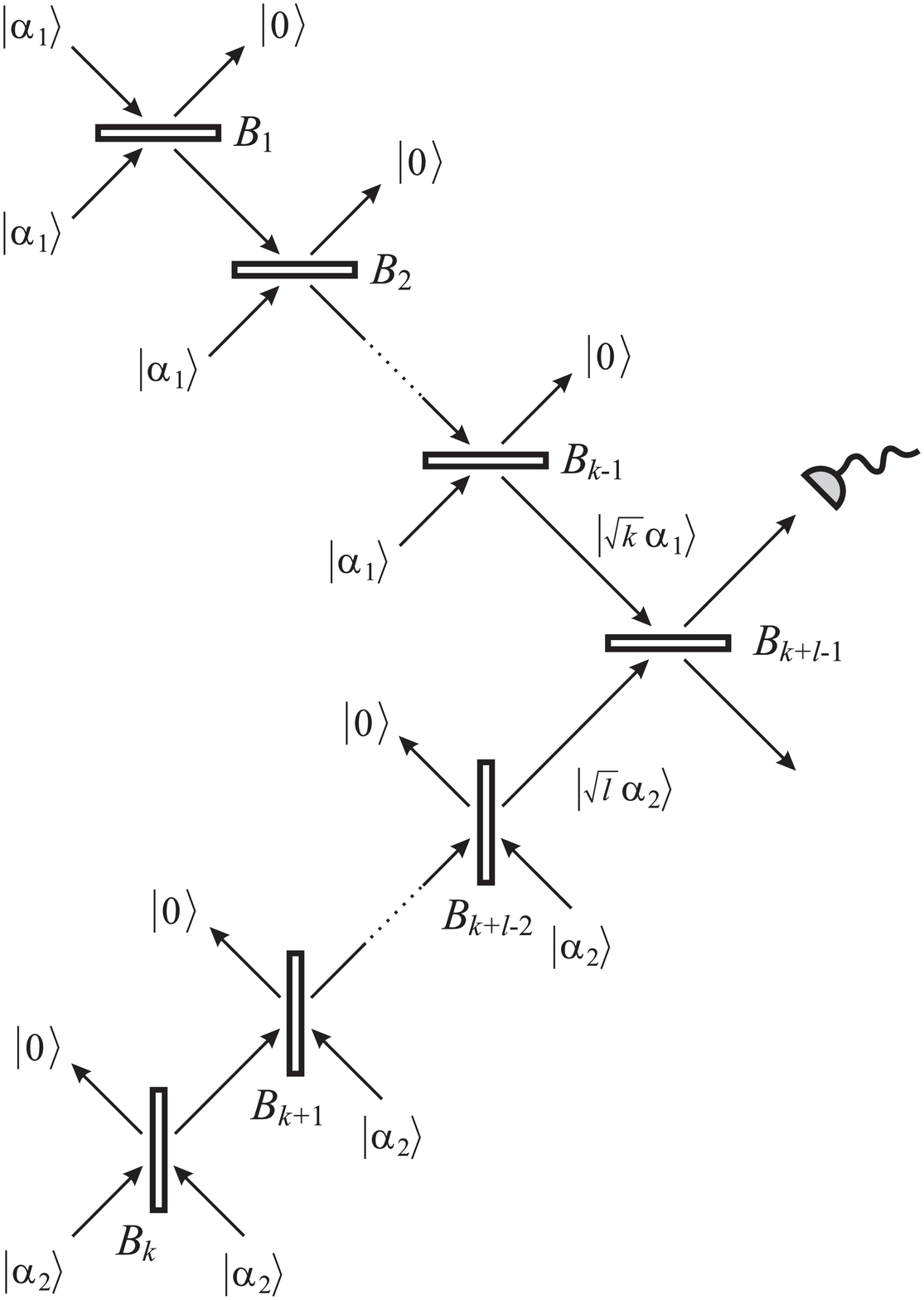}
\caption{The beam-splitter setup for the comparison of two finite-size ensembles composed of
$k$ copies of the coherent  state $\ket{\alpha_1}$  and $l$ copies of the  coherent  state $\ket{\alpha_2}$, respectively.}
\label{csetup}
\end{center}
\end{figure}

Our setup is composed of $k+l-1$ beam-splitters and one photodetector.
The $k-1$ beam-splitters are used to ``concentrate'' (focus) the information encoded
in $k$ copies of the first state. Namely, they are arranged
according to Fig.~\ref{csetup} and they perform the unitary transformation
$\ket{\alpha_1}^{\otimes k}\mapsto
\ket{\sqrt{k}\alpha_1}\otimes\ket{0}^{\otimes k-1}$. To do this
the transmissivities of the beam-splitters must be set as follows
\begin{eqnarray}
T_j=\frac{j}{j+1}\qquad R_j=\frac{1}{j+1}\, . \nonumber
\end{eqnarray}
Similarly, $l-1$ beam-splitters are used to ``concentrate'' the $l$ copies
of the second state. The ``concentrated'' states
$\ket{\sqrt{k}\alpha_1}$, $\ket{\sqrt{l}\alpha_2}$ are then launched into
the last beam-splitter in which the comparison of input coherent states is performed.
It performs the following unitary transformation
\begin{eqnarray}
\ket{\sqrt{k}\alpha_1}\otimes\ket{\sqrt{l}\alpha_2}&\mapsto&\ket{\sqrt{T_f k}\alpha_1+\sqrt{R_f l}\alpha_2} \nonumber\\
&\ &\otimes\ket{\sqrt{T_f l}\alpha_2-\sqrt{R_f k}\alpha_1}\, ,
\end{eqnarray}
where $R_f,T_f$ is the reflectivity and transmissivity of the last
beam-splitter. To obtain the vacuum in the upper output (see Fig.3)
we need to adjust the values of reflectivity and transmissivity so that
the identity $kR_f=lT_f$ holds, i.e.
\begin{eqnarray}
T_f=\frac{k}{k+l}, \qquad R_f=\frac{l}{k+l}. \nonumber
\end{eqnarray}

Finally, a photodetector will measure the presence of photons in the
upper output port of the last beam-splitter (see Fig.~\ref{csetup}).
If the two compared states are identical, in the output port
we have zero photons - that is this port is in the vacuum state. Therefore
a detection of at least one photon  unambiguously indicates the difference
between the compared states.
On the other hand the observation of no photons is inconclusive,
since each coherent state has a nonzero overlap with the vacuum.
As a result we obtain the success probability
\begin{eqnarray}
P(\ket{\alpha_1},\ket{\alpha_2})&=&1-|\Bra{0}\ket{\sqrt{\frac{kl}{k+l}}
(\alpha_2-\alpha_1)}|^2 \nonumber\\
&=&1-e^{-\frac{kl}{k+l}|\alpha_1-\alpha_2|^2}\, ,
\label{prbs}
\end{eqnarray}
which is the optimal one.
Analyzing the last equation we find out that
$P(\ket{\alpha_1},\ket{\alpha_2},m,n)>P(\ket{\alpha_1},\ket{\alpha_2},k,l)$
if and only if $\frac{mn}{m+n}>\frac{kl}{k+l}$.
This equivalence implies that
$P(\ket{\alpha_1},\ket{\alpha_2},k+1,l)>P(\ket{\alpha_1},\ket{\alpha_2},k,l)$.
Thus, also in the case of coherent states the additional copy of one
of the compared states helps to increase the mean success of the state comparison. For a fixed number of copies
of both compared states $N$ the fraction $k(N-k)/N$ is maximized for $k=N/2$.
Therefore, the probability of revealing the difference of the states
is maximized if $k=l$.

\section{Conclusion}
Let us summarize our main results on the quantum-state comparison derived in this paper.
The difference of the unknown states $\ket{\psi_1},\ket{\psi_2}$ can be
unambiguously detected with the success rate
\begin{eqnarray}
P(\ket{\psi_1},\ket{\psi_2})=1-
\sum_{m=0}^{\min(k,l)}\frac{\binom{k}{m}\binom{l}{m}}{\binom{k+l}{k}}
|\Bra{\psi_1}\ket{\psi_2}|^{2m}\, ,
\end{eqnarray}
providing that we have $k$ copies of state $\ket{\psi_1}$
and $l$ copies of the state $\ket{\psi_2}$. This result does not depend
on the dimension of the system in contrast to the average success rate,
which reads
\begin{eqnarray}
\overline{P(k,l)}=1-\frac{\dim(\Hs^{\otimes k+l}_{sym})}{\dim(\Hs^{\otimes k}_{sym})\dim(\Hs^{\otimes l}_{sym})}\, .
\end{eqnarray}
Given the a priori
knowledge that the states are coherent one can increase the
probability (see Fig.1) to
\begin{eqnarray}
P(\ket{\alpha_1},\ket{\alpha_2})=1-e^{-\frac{kl}{k+l}|\alpha_1-\alpha_2|^2}\, .
\end{eqnarray}
The improvement is significant (Fig.1) for
small number of copies.

We also addressed the problem of maximizing the success probability
providing that the total number of available copies is fixed. We have shown
that it is optimal if the number of copies
is the same, i.e. $k=l=N/2$. In the limit of the large number of copies
the comparison approach reduces to ``classical'' comparison based
on the quantum-state estimation.

We have proposed an optical implementation of the optimal quantum-state
comparator of two finite ensembles of coherent states.
This proposal is relatively easy to implement, since it
consists only of $N-1$ beam-splitters and a single photodetector. Unfortunately,
the  success of unambiguous state comparison is very fragile with respect to
small imperfections. The reason is that the device can be only used for 
pure states. Therefore our device can be used only in the situation 
when sources of a noise $\cal N$ can be modeled as quantum channels 
preserving the validity of the no-error conditions
${\rm Tr}(\Pi_1^{\rm coh}{\cal N}[\Delta_{\rm coh}^N])=0$. An example of such
noise is an application of random unitary channel (simultaneously on all copies)
transforming coherent states into coherent states.

\section*{ACKNOWLEDGMENTS}
This work was supported by the European Union projects QAP,
CONQUEST, by the Slovak Academy of Sciences via the project CE-PI,
and by the projects APVT-99-012304, and VEGA. Authors want to thank Teiko Heinosaari  for  helpful discussions.


\appendix

\section{Evaluation of $\overline{P(k,l)}$ }
Before calculating the average of $P(\ket{\psi_1},\ket{\psi_2})$ it is useful
to evaluate the mean values of the overlaps
\begin{eqnarray}
& & \overline{|\Bra{\psi_1}\ket{\psi_2}|^{2m}}=
\int_{S_d}\int_{S_d}d\psi_1 d\psi_2
\Bra{\psi_1}\ket{\psi_2}^m\Bra{\psi_2}\ket{\psi_1}^m \nonumber
\\ & &
=\int_{S_d}d\psi_1 \bra{\psi_1}^{\otimes m}
\left ( \int_{S_d} d\psi_2 \ket{\psi_2}\bra{\psi_2}^{\otimes m} \right )
\ket{\psi_1}^{\otimes m} \nonumber
\\ & &
=\frac{1}{\binom{m+d-1}{d-1}}\int_{S_d}d\psi_1
\bra{\psi_1}^{\otimes m}\sym \ket{\psi_1}^{\otimes m} \nonumber
\\ & &
=\frac{1}{\binom{m+d-1}{d-1}}\, , \label{movlp}
\end{eqnarray}
where we exploited the identity in Eq. (\ref{delta}).

We will insert Eqs. (\ref{prpure}) and (\ref{movlp})
into the definition (\ref{mpr}) and utilize the Vandermonde's identity
\begin{eqnarray}
\binom{a+b}{r}=\sum_{m=0}^{r}\binom{a}{m}\binom{b}{r-m} \nonumber
\end{eqnarray}
to evaluate the summation to obtain
\begin{eqnarray}
\overline{P(k,l)}&=&1-\frac{1}{\binom{k+l}{k}}\sum_{m=0}^{\min(k,l)}\frac{\binom{k}{m}\binom{l}{m}}{\binom{m+d-1}{d-1}}\nonumber\\
&=&1-\frac{k!(d-1)!}{(k+d-1)!}\frac{1}{\binom{k+l}{k}}\sum_{m=0}^{k}\binom{k+d-1}{k-m}\binom{l}{m}\nonumber\\
&=&1-\frac{k!(d-1)!}{(k+d-1)!}\frac{\binom{k+l+d-1}{k}}{\binom{k+l}{k}}\nonumber\\
&=&1-\frac{\binom{k+l+d-1}{k+l}}{\binom{k+d-1}{k}\binom{l+d-1}{l}}\nonumber\, .
\end{eqnarray}
The previous steps are valid for $k<l$, however we can perform analogous calculation for $l\leq k$ and obtain the same final result.

\section{Proof of lemma}
\leftline{\bf Lemma}
\noindent
Suppose we have a polynomial $Q_r(x)=\sum_{m=0}^{r}a_m x^m$ with the
following properties:
\begin{enumerate}
\item{}$Q_r(1)=0$
\item{}$a_m\geq 0$ for $m\leq r_0$ and $a_m\leq 0$ for $r_0<m\le r$
\end{enumerate}
Then $Q_r(x)\geq 0$ for all $x\in [0,1]$.

\noindent{\it Proof:}
For $x\in [0,1]$ and $a>b$ it follows that $x^a<x^b$. Therefore we
can write
\begin{eqnarray}
\nonumber
Q_r(x)&=&\sum_{m=0}^{r_0}a_m  x^m+\sum_{m=r_0 +1}^{r}a_m  x^m \\
&\geq& x^{r_0}\sum_{m=0}^{r_0}a_m+x^{r_0 +1}\sum_{m=r_0 +1}^{r}a_m\\
&=&(1-x)x^{r_0}\sum_{m=0}^{r_0}a_m \\
&\geq& 0\, ,
\end{eqnarray}
where we have used the fact that
$0=Q_r(1)=\sum_{m=0}^{r_0}a_m+\sum_{m=r_0 +1}^{r}a_m$, i.e.
$\sum_{m=r_0 +1}^{r}a_m=-\sum_{0}^{r_0}a_m$.

\section{Projectors onto coherent states}
Coherent states $\ket{\alpha}$ are intimately related to the group
of phase-space displacements $G$ generated by the Glauber operator
$D_\alpha=\exp(\alpha a^\dagger-\alpha^* a)$ via
the following relation $D_\alpha\ket{0}=\ket{\alpha}$, where $\ket{0}$
is the vacuum (ground) state of a harmonic oscillator.
Using the group invariant measure $d g$ (its support contains all
coherent states) the operator $\Delta$ can be expressed as follows
\begin{equation}
\Delta = \int_G dg (D_g\ket{0}\bra{0}D_g^\dagger)^{\otimes N}.
\end{equation}
Applying the theorem proved in Ref.~\cite{shucker} to the representation
of the group of displacements we find that
\begin{equation}
\Delta=\int_G dg (D_g\ket{0}\bra{0} D_g^\dagger)^{\otimes N}
=\lambda\Delta_{\rm coh}^{N}\, ,
\end{equation}
where $\lambda$ is a positive number ($\Delta$ is positive)
and $\Delta^{N}_{\rm coh}$ is the projector onto the linear subspace
spanned by the product states $\ket{\alpha}^{\otimes n}$.
A particular choice of the group invariant measure $dg$ affects
the value of the parameter $\lambda$. Our goal is to calculate
the projector $\Delta_{\rm coh}^N$, hence we are looking for a measure
$dg$ such that $\lambda=1$. The canonical Lebesgue measure $d\alpha$
on the complex plane $\complex$ is invariant under
complex translations (displacements) and therefore the correct
measure $dg$ is proportional to
$d\alpha$, that is  $dg=\mu d\alpha$ for some positive number $\mu$, i.e.
\begin{equation}
\Delta_{\rm coh}^N=\mu\int_{\complex} d\alpha
\ket{\alpha}\bra{\alpha}^{\otimes N} \, .
\end{equation}

Now, setting $\alpha = r e^{i\theta}$, we have, expanding the coherent states in terms of number
states,
\begin{eqnarray}
\nonumber
\Delta_{\rm coh}^N |0\rangle^{\otimes N} & = &
\mu\int_\complex d\alpha e^{-N|\alpha |^{2}/2} \times \\
\nonumber & & \times
\sum_{l_{1}=0}^{\infty}
\frac{\alpha^{l_{1}}}{\sqrt{l_{1}!}} \ldots \sum_{l_{N}=0}^{\infty} \frac{\alpha^{l_{N}}}{\sqrt{l_{N}!}}
(\langle \alpha |0\rangle )^{N} |l_{1},\ldots l_{N}\rangle \nonumber \\
\nonumber
& = & 2\pi\mu
\int_{0}^{\infty} dr\ r e^{-Nr^{2}} |0\rangle^{\otimes N}  \nonumber \\
 & = & \mu\frac{\pi}{N} |0\rangle^{\otimes N} \, ,
\end{eqnarray}
because $\int_0^{2\pi} e^{i\theta(l_1+\dots +l_N)}d\theta=2\pi$
if $l_1+\dots +l_N=0$, and vanishes otherwise.
The invariance of the canonical Lebesgue measure implies that
\begin{eqnarray}
\Delta_{\rm coh}^N D_\beta^{\otimes N} &=& D_\beta^{\otimes N} D_{-\beta}^{\otimes N}\Delta_{\rm coh}^N D_\beta^{\otimes N} \nonumber\\
&=&D_\beta^{\otimes N}\mu\int_{\complex} d\alpha \ket{\alpha-\beta}\bra{\alpha-\beta}^{\otimes N}\nonumber\\
&=&D_\beta^{\otimes N}\mu\int_{\complex} d(\alpha-\beta) \ket{\alpha-\beta}\bra{\alpha-\beta}^{\otimes N}\nonumber\\
&=&D_\beta^{\otimes N}\mu\int_{\complex} d\alpha \ket{\alpha}\bra{\alpha}^{\otimes N}\nonumber\\
&=&D_\beta^{\otimes N} \Delta_{\rm coh}^N
\label{appc6}
\end{eqnarray}
The previous identity (\ref{appc6}) implies
\begin{equation}
 \Delta_{\rm coh}^N |\beta\rangle^{\otimes n}
= \Delta_{\rm coh}^N D_\beta^{\otimes N} |0\rangle^{\otimes N} =
D_\beta^{\otimes N} \Delta_{\rm coh}^N |0\rangle^{\otimes N} \, .
\end{equation}
Consequently, for all $\ket{\psi}\in{\cal H}_{\rm coh}\equiv
{\rm span}\{\ket{\alpha}^{\otimes N}\}$ it holds that
\begin{equation}
\Delta_{\rm coh}^N |\psi\rangle = \mu\frac{\pi}{N} |\psi\rangle \, ,
\end{equation}
and for all $|\psi_\perp\rangle\in\mathcal{H}_{0}^{\perp}$
we have $\Delta_{\rm coh}^N |\psi_\perp\rangle = 0$. The above
equality fixes the invariant measure $dg$ to be $\frac{N}{\pi}d\alpha$,
where $d\alpha$ is the Lebesgue measure on the complex plane.


\end{document}